\documentstyle[multicol,aps,epsfig,amstex]{revtex} 
 
\tighten       
 
\begin{document} 
\draft 
 
\title{Theory of transmission through disordered superlattices} 
\author{Andreas Wacker} 
\address{ Institut f{\"u}r Theoretische Physik,  
Technische Universit{\"a}t Berlin,  
Hardenbergstr. 36, 10623 Berlin, Germany} 
\author{Ben Yu-Kuang Hu} 
\address{Mikroelektronik Centret, 
Danmarks Tekniske Universitet, DK-2800 Lyngby, Denmark and\\ 
Department of Physics, 250 Buchtel Commons, University of Akron, 
Akron, OH 44325-4001\footnote[1]{Present and permanent address}}

\date{final version, accepted by Physical Review B} 
\maketitle 
 
\begin{abstract} 
We derive a theory for transmission through disordered finite superlattices 
in which the interface roughness scattering is treated by disorder  
averaging.  This procedure permits efficient calculation of the transmission through  
samples with large cross-sections.  These calculations can be   
performed utilizing either the Keldysh or the Landauer-B\"uttiker  
transmission formalisms, both of which yield identical equations. 
For energies close to the lowest miniband, we demonstrate the accuracy of 
the computationally efficient Wannier-function approximation. 
Our calculations indicate that the transmission is strongly affected 
by interface roughness and that information about 
scale and size of the imperfections can be obtained from transmission 
data. 
\end{abstract} 
 
\begin{multicols}{2}

\section{Introduction} 
 
Semiconductor superlattices continue  
to attract substantial interest both among fundamental and applied  
researchers.  
One motivating factor is the possibility of tailoring the miniband  
structure\cite{ESA70,SHI75,GRA95a} for device purposes.  Furthermore,  
a large variety of other physical phenomena such as the formation of  
Wannier-Stark ladders \cite{MEN90}, negative differential 
conductance \cite{SIB90}, and Bloch oscillations\cite{LEO96} 
can be observed in superlattices.   
 
The presence of minibands has been probed directly  
by investigating the transmission of ballistic electrons   
through short semiconductor superlattices \cite{KUA92,RAU97}.  
In recent experiments the quenching of the miniband structure 
by an applied electric field was also demonstrated \cite{RAU97a}. 
Comparison of further experiments with theoretical  
calculations indicated a strong influence of scattering on the transmission, 
and it was argued that interface roughness might cause significant 
deviations from pure ballistic transmission through the  
sample\cite{RAU98,WAC99a}. 
 
A good understanding of the transmission characteristics through  
short superlattices is important as these structures are used as energy filters. 
For example, in quantum cascade lasers, superlattice filters are used to 
selectively populate the upper energy level of the active region \cite{CAP97}. 
The most straightforward way to calculate the transmission through 
a superlattice is the transfer matrix method \cite{TSU73}. 
Alternatively, the Schr{\"o}dinger equation of 
the superlattice can be solved directly. 
These methods typically assume homogeneity in the direction 
perpendicular [the $(x,y)$-plane] to the superlattice.  The momenta 
in the $(x,y)$-plane are then good quantum numbers and decouple from the  
superlattice direction, reducing the problem to a one-dimensional calculation;  
see, {\it e.g.}, Ref.~\cite{GLY91}.  The one-dimensional calculation  
can handle fluctuations \cite{DIE94} in the well or barrier thickness. 
However, real samples also exhibit 
a lack of periodicity in the $(x,y)$-plane due to the presence of  
impurities and interface roughness.  This can change the transport properties 
essentially, as the states with different parallel momenta couple 
to each other.  This $(x,y)$-plane inhomogeneity can be tackled by solving the  
Schr{\"o}dinger equation on a mesh for the full three dimensional 
structure \cite{TIN95}.  Alternatively, the method of Green-functions, 
based on Ref.~\cite{CAR71}, may be used (see Ref.~\cite{DAT95}  
for an easily accessible presentation of the method). Recently such an approach 
has been presented for a full calculation of the current through 
a resonant tunneling diode where both interface roughness and 
phonon scattering have been taken into account \cite{LAK97}. 
However, these simulations use a fine grid and are hence unsuitable  
for longer structures such as superlattices consisting of many wells,  
since the number of grid points increases dramatically. 
 
In this paper we propose a new method for such calculations which 
significantly reduces the computational complexity.  
We treat the inhomogeneity in the $(x,y)$-plane by averaging over 
disorder configurations.  The number of grid points  
in the $z$-direction is reduced significantly by restricting to the basis set to the  
Wannier functions localized in the wells.  This Wannier approximation is shown  
to be valid near the resonance condition if the energy gap between the minibands 
is large compared to the bias and the widths of the minibands 
themselves.  We compare our results to calculations on a finite grid 
and find good agreement.  Our method has the advantage that it corresponds  
to infinitely large cross-sections and hence, unlike the finite-grid 
calculations, does not show configuration dependent fluctuations. 
 
The paper is  organized as follows. 
We present the general model within which 
our calculations are performed in section \ref{SecGenformal}.  In section  
\ref{SecApplication}, we describe the 
approximations which allow us to perform practical calculations of extended 
superlattice structures. Our results are presented in section \ref{Secresults} 
and we conclude with a summary.  Appendix \ref{SecImpurityav}  
shows the equivalence of the Landauer-B\"uttiker transmission formalism  
with the approach by 
nonequilibrium Green functions for the case of impurity averaging.  
In appendix \ref{SecAccuracy} 
we justify the approximations used in section \ref{SecApplication}. 
As many different symbols appear in this paper, for easy reference 
we display the frequently used ones in Table 1. 
 
\section{General formalism} 
\label{SecGenformal} 
 
In this paper, we study transport through a superlattice contacted 
to external voltage sources via leads.  
We model the superlattice as an active central region  
coupled to noninteracting lead regions.  This is the general  
approach described in Refs.~\onlinecite{CAR71,DAT95,MEI92,JAU94} and  
\onlinecite{HAU96}. 
In this section we briefly review this approach, introduce our notation, 
and discuss the issue of impurity averaging. 
 
We divide the sample into a central region C and  
lead regions, indexed by $\ell$. The Hamiltonian is 
\begin{equation} 
\hat H = \hat H_C + \sum_{\ell} \hat H_\ell +  
\hat H_{{\ell}C} + \hat H_{{\ell}C}^\dagger  
\end{equation} 
Here $\hat H_C$ and $\hat H_\ell$ are the terms for the 
central structure and leads, respectively, and  
$\hat H_{{\ell}C}$  is the coupling term from the center to the lead 
$\ell$. 
In this paper, we ignore electron--electron interactions beyond Hartree 
so all the above terms are single-particle-like.   
 
The central structure has states with the wave function  
$\phi_{C,j}(\vec{r})$, where $j$ is the eigenstate index. 
We assume that each lead $\ell$ is disorder-free so that the eigenstates  
can be separated into transverse and longitudinal parts,  
$\phi_{{\ell}\alpha q}(\vec{r})=\chi_{{\ell}\alpha}({\bf r}) 
\varphi^{\ell}_{q}(z)$, 
where $z$ is the spatial coordinate in the direction towards the central  
structure and ${\bf r}$ is a two-dimensional vector perpendicular to $z$. 
The index $\alpha$ numbers the modes within a given lead. The index  
$q$ denotes the behavior far away from the central region where 
$\varphi^{\ell}_q(z) \sim e^{iqz}$ is assumed.

\subsection{Green functions and current through structure} 
 
The current through a structure can be determined by  
the Green function of the structure {\sl in the presence of 
coupling to the leads}, given by a matrix {\bf G} with matrix elements  
\begin{mathletters} 
\begin{eqnarray} 
G^{<}_{ij}(t,t')&=& i\langle \hat{c}_j^\dagger(t') \hat{c}_i(t)\rangle\\ 
G^{\rm ret\atop adv}_{ij}(t,t') &=& \mp i \langle\{\hat{c}_i(t), 
\hat c_j^\dagger(t')\}\rangle\ \theta(\pm (t-t')\;) . 
\end{eqnarray} 
\end{mathletters} 
Here $\hat{c}_i^\dagger\;(\hat{c}_i)$ are fermion creation  
(annihilation) operators of states $\phi_{C,i}$ in the central 
region, and $\{\cdots,\cdots\}$ denote anticommutators. 
 
In the following we consider time-independent problems, so that 
${\bf G}$ only depends on $t-t'$, and we work in energy-space by Fourier  
transforming ${\bf G}$ with respect to $t-t'$.  
The net current from mode $\alpha$ in lead $\ell$ into the structure  
is given by\cite{MEI92,HAU96} 
\begin{equation} \begin{split}
J_{\ell\alpha} =2 \frac{ie}{\hbar}\int\frac{dE}{2\pi} 
{\rm Tr}\big\{&\boldsymbol{\Gamma}_{\ell\alpha}(E) \left[{\bf G}^{<}(E)\right.\\ 
&\left. +f_{\ell\alpha}(E)\left({\bf G}^{\rm ret}(E)-{\bf G}^{\rm adv}(E)\right) 
\right]\big\} \, .\label{EqCurrentalpha} 
\end{split}\end{equation} 
Here $f_{\ell\alpha}(E)$ gives the occupation of a state with  
energy $E$ in lead $\ell$ for the mode $\alpha$,  
$e<0$ is the charge of the electron, and $\boldsymbol{\Gamma}_{l,\alpha}$  
is a parameter 
describing the coupling between the states in the central region and the 
leads [see Eq.\ (\ref{Eqdefine-Gamma}) below].  The factor of 2 is for 
spin.  
 
To describe transmission through the superlattice, we need to  
obtain expressions for the right hand side of 
Eq.\ (\ref{EqCurrentalpha}).  We do so as follows. 
We first define $\hat H_{C,0}$ and $\hat H_{C}'$,  
as the ordered, solvable part and the disordered part of the central 
region Hamiltonian, respectively, and $\hat H_C = \hat H_{C,0} + \hat H_{C}'$.   
The retarded Green function for the structure is determined by the equation 
(see Ref.~\cite{HAU96}, chapter 12)  
\begin{equation} 
\left( E-{\bf H}_{C,0}-\boldsymbol{\Sigma}_{C}^{\rm ret} -  
\sum_{\ell\alpha}\boldsymbol{\Sigma}^{\rm ret}_{\ell\alpha}(E)\right) 
{\bf G}^{\rm ret}(E)=\boldsymbol{1}\,. \label{EqGret} 
\end{equation} 
The term $\boldsymbol{\Sigma}^{\rm ret}_C$ is the irreducible self-energy  
due to $\hat H_{C}'$.  In cases where $\hat H_C'$ contains interparticle  
interactions, $\boldsymbol{\Sigma}^{\rm ret}_C$ is often very difficult to calculate;  
however, for static disorder, simply  
$\boldsymbol{\Sigma}^{\rm ret}_C = {\bf H}_C'$. 
The term $\boldsymbol{\Sigma}^{\rm ret}_{\ell\alpha}(E)$  
gives the self-energy contributions due to the coupling of the central region 
to lead $\ell$ and mode $\alpha$,  
\begin{eqnarray} 
\Sigma^{\rm ret}_{\ell\alpha,ij}(E)&=& 
\sum_q \langle \phi_{C,i}|\hat{H}_{\ell C}^\dagger|\phi_{\ell\alpha q}\rangle  
\langle \phi_{\ell\alpha q}|\hat{H}_{\ell C}|\phi_{C,j}\rangle\; 
g^{\rm ret}_{\ell\alpha q}(E)\nonumber \\ 
&=&\frac{L_{\ell}}{2\pi}\int_0^{\infty} dE_q\ \frac{2}{\hbar v_q} 
\langle \phi_{C,i}|\hat{H}_{\ell C}^\dagger|\phi_{\ell\alpha q}\rangle 
\nonumber \\  
&&\phantom{\frac{L_{\ell}}{2\pi}\int_0^{\infty} dE_q} 
\langle \phi_{\ell\alpha q}|\hat{H}_{\ell C}|\phi_{C,j}\rangle\; 
g^{\rm ret}_{\ell\alpha q}(E) 
\label{Eqsigmaalpha} 
\end{eqnarray} 
where we have taken the continuum limit  
$\sum_q\to L_{\ell}/2\pi\int_{-\infty}^{\infty} dq$ ($L_\ell$ is the 
length of lead $\ell$). 
The factor $2$ results from the two possible  
values $\pm q$ for a given energy $E_q$,  
$g^{\rm ret}_{\ell\alpha q}(E)=1/(E-E_q-E_{\ell\alpha}+i0^+)$ is the 
free-particle Green function of the lead in absence of the central 
region, $E_q=\hbar^2q^2/2m^*$, $v_q=\hbar q/m^*$ and 
$E_{{\ell},\alpha}$ is the lateral energy of the mode $\alpha$. Here $m^*$ 
is the effective electron mass. 
Note that $G_{ij}^{\rm adv}(E) = [G_{ji}^{\rm ret}(E)]^*$, since 
we have a time-independent system. 
 
The coupling parameter $\boldsymbol{\Gamma}_{\ell\alpha}$ is defined by  
\begin{eqnarray} 
{\Gamma}_{\ell\alpha,ij}(E) 
&=&i\left[{\Sigma}^{\rm ret}_{\ell\alpha,ij}(E)- 
{\Sigma}_{\ell\alpha,ij}^{\rm adv}(E)\right]\nonumber \\ 
&\equiv&\frac{2 L_{\ell}\langle \phi_{C,i}|\hat{H}^\dagger_{\ell C}| 
\phi_{\ell\alpha\; q(E-E_{\ell\alpha})}\rangle}
 {\hbar v_{q(E-E_{\ell\alpha})}}\nonumber\\
&&\times
\langle \phi_{\ell\alpha\; q(E-E_{\ell\alpha})}|
\hat{H}_{\ell C}|\phi_{C,j}\rangle \; 
\Theta(E-E_{{\ell}\alpha}) 
\label{Eqdefine-Gamma} 
\end{eqnarray} 
where $q({\cal E})=\sqrt{2m^*{\cal E}}/\hbar$. $\Theta(x)$ is the  
Heavyside function with $\Theta(x)=1$ for $x\ge 0$ and  
$\Theta(x)=0$ for $x< 0$. 
Note that $\Gamma_{\ell\alpha,ij}(E)=0$ for $E<E_{\ell\alpha}$  
since there are no propagating states into 
which the central-region states can tunnel. 
 
${\bf G}^{<}(E)$ can be obtained by the Keldysh relation\cite{KEL65} 
\begin{equation} 
{\bf G}^{<}(E)={\bf G}^{\rm ret}(E) 
\boldsymbol{\Sigma}^<(E){\bf G}^{\rm adv}(E)\label{EqKeldysh} 
\end{equation} 
where 
\begin{equation} 
\boldsymbol{\Sigma}^<(E)=\boldsymbol{\Sigma}^{<}_{C}(E)+\sum_{{\ell}\alpha} 
\boldsymbol{\Sigma}_{\ell\alpha}^{<}(E). 
\label{EqSigmaless} 
\end{equation} 
Here, $\boldsymbol\Sigma_C^<$ is the self-energy resulting from 
scattering inside the  
structure.  For a fixed disorder potential, this term is identically zero. 
The term $\boldsymbol\Sigma_{\ell\alpha}^<(E)$ is the self-energy due to 
the presence of the coupling to the leads, 
\begin{eqnarray} 
\Sigma_{\ell\alpha,ij}^<(E)&=& 
\sum_q \langle \phi_{C,i}|\hat{H}_{\ell C}^\dagger|\phi_{\ell\alpha q}\rangle  
\langle \phi_{\ell\alpha q}|\hat{H}_{{\ell}C}|\phi_{C,j}\rangle  
g^{<}_{\ell\alpha q}(E)\nonumber \\ 
&=&i\,\Gamma_{\ell\alpha,ij}(E)\, f_{\ell\alpha}(E)\label{EqSigmalessL} 
\end{eqnarray} 
where we have used $g^{<}_{\ell\alpha q}(E)= 
-2if_{\ell\alpha}(E)\,{\rm Im}\left\{g^{\rm ret}_{\ell\alpha q}(E)\right\}$. 
The occupation function $f_{\ell\alpha}(E)$ in lead $\ell$ 
and is given by the externally imposed conditions.  
Usually, the leads are assumed to be in  
thermal equilibrium and hence a Fermi distribution with chemical potential  
$\mu_{\ell}$, independent of $\alpha$, is used. In contrast, the  
different modes can be populated individually by injection, as discussed 
later, so that we want to keep the full function $f_{\ell\alpha}(E)$. 
 
\subsection{Relation to the Landauer-B{\"u}ttiker approach} 
 
The Landauer-B\"uttiker approach has been used extensively to 
study transmission through mesoscopic structures, and consequently 
many people are familiar with the formalism. 
As the Keldysh formulation is not as widely known, 
in this subsection we demonstrate the equivalence of the  
two approaches for transport through a system with static disorder. 
 
The retarded and advanced Green functions can be expressed in terms 
of $\Gamma$ via 
\begin{equation} 
{\bf G}^{\rm ret}(E)-{\bf G}^{\rm adv}(E)=-i{\bf G}^{\rm ret}(E) 
\boldsymbol{\Gamma}(E){\bf G}^{\rm adv}(E)\label{Eqspectral} 
\end{equation} 
where the total scattering rate $\boldsymbol{\Gamma}$ has two contributions 
\begin{equation} 
\boldsymbol{\Gamma}(E)= 
i\left[\boldsymbol{\Sigma}^{{\rm ret}}_{C}(E) 
-\boldsymbol{\Sigma}^{{\rm adv}}_{C}(E)\right]+ 
\sum_{\ell\alpha}\boldsymbol{\Gamma}_{\ell\alpha}(E) 
\end{equation} 
resulting from scattering inside the structure and transitions into the leads. 
If the scattering within the structure itself is purely elastic and 
is treated in a particular fixed configuration as a potential in 
Eq.~(\ref{EqGret}),  
then $\boldsymbol{\Sigma}^{\rm ret}_C = {\bf H}_C'$ 
and $\boldsymbol{\Sigma}^{<}_C(E)=0$; hence we may insert 
Eqs.~(\ref{EqKeldysh}) and (\ref{Eqspectral}) into 
Eq.~(\ref{EqCurrentalpha}) and find the Landauer-B{\"u}ttiker  
expression \cite{MEI92,BUE90} 
\begin{equation} 
J_{\ell\alpha}=2 \frac{e}{\hbar}\int\frac{dE}{2\pi} 
\sum_{\ell'\beta} T_{\ell\alpha\leftarrow \ell'\beta}(E)  
\left[f_{\ell\alpha}(E)-f_{\ell'\beta}(E)\right] 
\label{EqBuett} 
\end{equation} 
(factor of 2 for spin) with the transmission matrix 
\begin{equation} 
T_{\ell\alpha\leftarrow \ell'\beta}(E)= 
{\rm Tr}\left\{\boldsymbol{\Gamma}_{\ell\alpha}(E) 
{\bf G}^{\rm ret}(E) 
\boldsymbol{\Gamma}_{\ell'\beta}(E){\bf G}^{\rm adv}(E)\right\}. 
\label{Eqtransmat} 
\end{equation} 
(There are several alternate ways to derive this result; {\it e.g.}, 
Ref.~\cite{DAT95} uses spatial discretization.) Note that 
Eq.~(\ref{Eqtransmat}) does 
{\sl not} hold if the scattering process is inelastic or the 
elastic scattering by static disorder is 
described by a self-energy obtained by configuration averaging.
In both cases 
$\boldsymbol{\Sigma}^{<}_C(E)\neq 0$ in contrast to the assumption leading to
Eq.~(\ref{EqBuett}).
 
\subsection{Impurity Averaging} 
\label{SubsecImpav} 
Eq.\ (\ref{Eqtransmat}) is exact for a given configuration of impurities  
and roughness, {\it i.e.}, for a specific ${\bf H}_C'$.  However, 
obtaining the transmission by simulating individual configurations is not 
computationally efficient, and hence it is advantageous to average over  
impurity configurations. In particular, such a procedure reestablishes   
symmetries which are broken by specific impurity configurations, thus 
simplifying the calculation significantly. 
 
After impurity averaging, we obtain 
\begin{equation} 
\overline{\bf G}^{\rm ret}(E) =  
[E + i0^+ - {\bf H}_{C,0} - {\boldsymbol{\overline\Sigma}}_C^{\rm ret} 
(E) - \sum_{\ell} 
{\boldsymbol{\Sigma}}_{\ell}^{\rm ret}(E)]^{-1}, 
\label{EqGretav} 
\end{equation} 
where the overlines indicate averages over disorder configurations in  
the central region.  Note that the disorder averaging introduces 
non-zero self-energies $\overline{\boldsymbol{\Sigma}}_C^{\rm ret}(E)$ and 
$\overline{\boldsymbol{\Sigma}}_C^{<}(E)$. 
As $\overline{\boldsymbol{\Sigma}}_C^{<}(E)\neq 0$ one 
cannot simply use Eq.\ (\ref{Eqtransmat}) with the $G$'s replaced 
by $\overline{G}$.  
In order to describe configuration-averaged elastic scattering within the 
transmission formalism, the averaging must be performed for the total  
transmission matrix in Eq.~(\ref{Eqtransmat}), and not just on the 
individual $G^{\rm ret}$ and $G^{\rm adv}$. 
This procedure is similar to the calculation of bulk conductivities 
using the Kubo formula, where it is crucial to include vertex corrections  
which fulfill the Ward-identity (see, e.g., Ref.~\cite{DON74}). 
We perform such a calculation in appendix \ref{SecImpurityav} 
for the superlattice structure discussed in section \ref{SecApplication}. 
We use the self-consistent Born-approximation for the scattering 
and therefore the appropriate vertex function is the so-called ladder approximation. 
 
The application of the more general Keldysh approach to calculate the current in 
the configuration averaged case is more straightforward, in that one {\em can} replace 
the $G$ by $\overline{G}$ in Eq.~(\ref{EqCurrentalpha}).  
Therefore, in order to evaluate the current, 
we need $\overline{\bf G}^<(E)$ and $\overline{\bf G}^{\rm ret}(E)$. 
The general iterative procedure for computing these is as follows.\label{Secprocedure} 
First the self-energies $\boldsymbol{\Sigma}_{\ell}^{\rm ret}(E)$  
and $\boldsymbol{\Sigma}_{\ell}^<(E)$ due to the  
leads are evaluated by Eqs.~(\ref{Eqsigmaalpha})  
and (\ref{EqSigmalessL}).  As these terms are independent 
of disorder configuration,  
these need only be evaluated once and then stored. 
With $\boldsymbol{\overline{\Sigma}}_C^{{\rm ret},<}$ initially set equal to zero,  
$\overline{\bf G}^<$ and $\overline{\bf G}^{\rm ret}$ are calculated.   
These $\overline{\bf G}$'s are  
used to calculate the $\boldsymbol{\overline{\Sigma}}_C$'s,  
via an appropriate approximation scheme. 
The updated $\boldsymbol{\overline{\Sigma}}_C$'s are used to generate new  
$\overline{\bf G}$'s via  
Eqs.~(\ref{EqKeldysh}) and (\ref{EqGretav}), and the 
process is iterated until convergence is achieved. 
Finally, the current is evaluated with  
Eq.~(\ref{EqCurrentalpha}).  
 
In Appendix \ref{SecImpurityav} we show explicitly that the  
ladder approximation for the vertex function in the transmission formulation
yields the same equations as the Keldysh approach within the self-consistent 
Born approximation, demonstrating the equivalence of the two methods for 
impurity scattering.  Nevertheless the Keldysh approach 
seems to be conceptually easier as there is only one place within this formulation
where an approximation is made; {\em i.e.}, in the self-energy.  In contrast, with 
the transmission formalism, errors can occur if the vertex function does not fulfill 
the Ward identity, providing a pitfall to trap the uninitiated and unwary. 
 
\section{Application to a superlattice structure} 
\label{SecApplication} 
 
\begin{figure}  
\noindent 
\epsfig{file=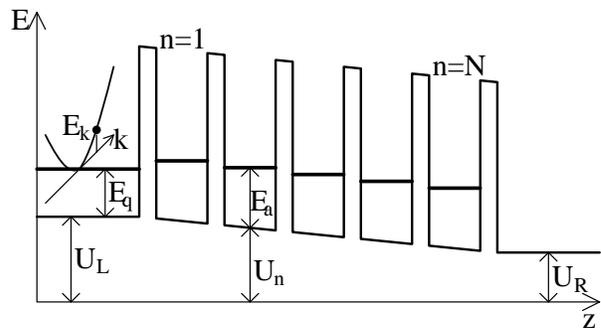,width=8cm}\\[0.2cm] 
\caption[a]{Sketch of the structure considered. 
The energy levels are indicated for $E_k=0$, which 
has to be added for finite parallel momentum $k$.} 
\label{Figstruktur} 
\end{figure}

Let us consider the superlattice structure sketched  
schematically in Fig.~\ref{Figstruktur}.  
The superlattice consists of $N$ identical wells 
embedded in $N+1$ barriers. A bias $U$ is applied to the structure 
yielding   constant 
potentials $U_L$ and $U_R=U_L+eU$ at the left and right contact,  
respectively. 
In order to perform calculations we now specify the 
basis states $\phi_{C,j}(\vec{r})$ and $\chi_{\ell\alpha}({\bf r})$ for our 
superlattice structure. 
The lead index $\ell$ takes two different values $L$ and $R$, 
for the left and right contact region, respectively. 
For superlattices with a large cross section $A$ it  
is natural to use a basis of plane waves  
$e^{i{\bf k}\cdot{\bf r}}/\sqrt{A}$ 
for the transverse coordinates $(x,y)$ both in the lead regions 
and in the superlattice itself. 
Then the index $\alpha$ of the 
states in the leads is 
replaced by ${\bf k}$ and we have 
$E_{(L/R,{\bf k})}=E_k+U_{L/R}$, where $E_k=\hbar^2k^2/2m$. 
 
\subsection{Wannier approximation for a superlattice\label{SecWA}} 
 
Let us now consider the central region; {\it i.e.,} the superlattice structure  
itself. 
In order to make a calculation tractable, we restrict ourselves 
to a subset of the basis functions of total Hilbert space,  
ignoring irrelevant high energy states.   
With respect to the $z$-direction inside the superlattice 
we use a basis of Wannier-functions 
$\Psi_n(z)$ ($n=1,\ldots N$) from the lowest miniband 
which are maximally localized in well $n$\cite{KOH59}. 
Such a basis has been successfully applied to  superlattice transport  
\cite{WAC98}. This approximation, which we call the Wannier 
approximation (WA), neglects higher mini-bands, and 
its validity is discussed in the Appendix \ref{SecAccuracy}. 
There we demonstrate that this approximation gives good results
for the transmission probability provided that the miniband width 
is smaller than the energy of the center of the miniband and 
the energy range of interest is sufficiently below the 
levels corresponding to the higher miniband states.
The states $\phi_{C,j}$ within the superlattice are, within the WA,  
given by products $\Psi_n(z)e^{i{\bf k}\cdot{\bf r}}/\sqrt{A}$, which can be 
labeled by $(n,{\bf k})$. 
 
Within the superlattice, the Green-function is determined by Eq.~(\ref{EqGret}) 
which in the WA basis reads 
\begin{equation} 
\begin{split} 
\sum_{n'{\bf k}'} 
\left[(E-E^a-E_k-U_n)\delta_{{\bf k},{\bf k}'}\delta_{n,n'} 
-H_{n{\bf k},n'{\bf k}'}\right.\\ 
-T_1\delta_{{\bf k},{\bf k}'}\left(\delta_{n,n'+1}+\delta_{n,n'-1}\right) 
\left.-\sum_{\ell\alpha}\Sigma^{\rm ret}_{\ell\alpha;\;n{\bf k},\,n'{\bf k}'}(E) 
\right]\\ 
G^{\rm ret}_{n'{\bf k}',m{\bf k}_1}(E)=\delta_{{\bf k},{\bf k}_1}\delta_{n,m} 
\label{EqGreenWannk} 
\end{split} 
\end{equation} 
Here, $U_n$ denotes the potential in the well $n$ (see Fig.~\ref{Figstruktur})
which is due to an external bias. 
(The mean-field potential induced by the carriers in the structure can be 
added as well.)
$T_1$ is the coupling between the wells and $E^a$ is the level energy of the 
Wannier state relative to the bottom of the well.   
For a given structure, we calculate $T_1$ and $E^a$ as follows. 
We consider first an infinite superlattice of the same composition. 
The eigenstates  in the infinite superlattice are Bloch functions with the 
miniband dispersion  $E^a(q)$. $E^a$ is then identified as the center of the 
miniband 
$d/(2\pi)\int dq E^a(q)$ and $T_1= d/(2\pi) \int dq E^a(q)\cos(qd)$, where 
$d$ is the period of the superlattice \cite{WAC98}; {\it i.e.}, $|T_1|$ 
is about a  fourth of the miniband width. 
Finally, $H_{n{\bf k},n'{\bf k}'}$ 
is the disorder scattering matrix element. 
 
If we average over disorder configurations the  
translational invariance in the $(x,y)$-plane is restored, 
and consequently, all impurity-averaged 
quantities are diagonal in ${\bf k}$ parallel to the $(x,y)$-plane. 
Therefore we are able to use the notation  
$\overline{G}_{n{\bf k},\, m{\bf k}}(E) \equiv \overline{G}_{nm}({\bf k},E)$ 
and matrices ${\bf G}({\bf k},E)$ and  $\boldsymbol{\Sigma}({\bf k},E)$  
have the components $G_{nm}({\bf k},E)$ and  $\Sigma_{nm}({\bf k},E)$, 
respectively. 
 
\subsection{Estimating the coupling and wide band limit} 
\label{SecCoupling} 
The coupling with the mode ${\bf k}$ in the left contact yields, 
from Eq.~(\ref{Eqsigmaalpha}), the self energy 
\begin{equation}\begin{split} 
&\Sigma^{\rm ret}_{L{\bf k};\;n{\bf k}_1,\, n'{\bf k}_2}(E)= 
\delta_{n,1}\delta_{n',1} 
\delta_{{\bf k}_1,{\bf k}}\delta_{{\bf k}_2,{\bf k}} \\
&\phantom{\Sigma}\times \frac{1}{2\pi} \int_0^{\infty} dE_q\  
\frac{2L_L|V_q|^2}{\hbar v_q} 
\frac{1}{E-E_q-E_{\bf k}-U_L+i0^+} 
\label{EqSigmawannierk}  
\end{split}\end{equation}  
where $V_q=\langle \varphi_{q}^L(z)|\hat{H}_{LC}|\Psi_1(z)\rangle$  
is the $z$-dependent part  
of the matrix element for the coupling to the  leads. 
Here we neglect the coupling to the inner wells $(n\neq 1)$, which should 
be small. 
The right contact gives the same term except with replacements  
$\delta_{n,1}\rightarrow \delta_{n,N}$ and $U_L,L_L\rightarrow U_R,L_R$. 
 
If the transmission function is strongly  
determined by resonances, only a small energy range of $E\approx E^a+U_L+E_k$  
contributes to the transmission. In this range we neglect the  
$q$ dependence of the coupling and extend the lower limit of the 
integration in Eq.~(\ref{EqSigmawannierk}) to $-\infty$. 
Then we obtain for the left lead 
\begin{eqnarray} 
\Sigma^{\rm ret}_{L{\bf k};\; n{\bf k}_1,\,n',{\bf k}_2}(E)&=& 
\delta_{n,1}\delta_{n',1} 
\delta_{{\bf k}_1,{\bf k}}\delta_{{\bf k}_2,{\bf k}} 
 \frac{-i}{2} \Gamma_L 
\end{eqnarray}  
with  
\begin{equation} 
\Gamma_L=\frac{2L_L|V_{q(E^a)}|^2}{\hbar v_{q(E^a)}}. 
\end{equation} 
This approximation is often referred to as wide band limit. 
Note that this limit becomes problematic if the voltage drop 
across the first barrier becomes large, as this changes the 
relevant values of $E$ and it cannot be regarded as constant  
[see also Appendix \ref{SecAccuracy}]. 
 
Now we want to estimate the value of $|V_q|^2$. 
For $E_q\approx E^a$ the wavefunction $\varphi_{q}^L(z)$ in the left lead  
behaves like the Wannier function $\Psi_0(z)=\Psi_1(z+d)$ 
which is localized in a fictitious additional well on the left side
of the structure. 
Now $\varphi_{q}^L(z)$ is normalized to $L_L$ while the spatial extension of 
the  Wannier function is given by $w_{\rm eff}$, which should be slightly larger 
than the well width, as the function penetrates into the barriers. 
Therefore we may set $\varphi_{q}^L(z)\sim \sqrt{w_{\rm eff}/L_L}\Psi_0(z)$. 
Then we can estimate the matrix element 
\begin{equation} 
\langle \varphi_q^L|H|\Psi_1\rangle\approx  
\sqrt{\frac{w_{\rm eff}}{L_L}}\langle \Psi_0|H|\Psi_1\rangle= 
\sqrt{\frac{w_{\rm eff}}{L_L}}T_1, 
\end{equation} 
yielding 
\begin{equation} 
\Gamma_L\approx \frac{2w_{\rm eff}T_1^2}{\hbar v_{q(E^a)}}. 
\end{equation} 
For the right contact, $\Gamma_R$ is given by the same value. 
 
\subsection{Interface Roughness} 
For ideal structures the potential $H_{n{\bf k},n'{\bf k}'}$ 
in Eq.~(\ref{EqGreenWannk}) is zero due to the translational 
invariance within the $(x,y)$-plane. 
However, interface fluctuations leading to well 
width fluctuations $\xi_n({\bf r})$ in real samples 
break this translational invariance.  If interwell 
scattering and well-width correlations between
different wells can be neglected, 
the averaged square of the scattering matrix element is given  
by \cite{GOO85,RID98} 
\begin{equation} 
\langle |H_{n{\bf k}+{\bf p},n'{\bf k}}|^2\rangle= 
\frac{K^2}{A}S({\bf p}) \delta_{n,n'}
\end{equation} 
where $K$ is equal to the change of energy  
$dE^a/dw$ per well width fluctuation \cite{RefTES} and 
$S({\bf p})$ is the Fourier  
transformation of the well width correlation function  
$\langle \xi_n({\bf r})\,\xi_n({\bf r'})\rangle=f({\bf r}-{\bf r'})$ which 
is assumed to be independent of the well index. The theory can be
extended to accommodate interwell scattering and well-width 
correlations between different wells (which may result from a
repetition of the microscopic interface structure over several 
superlattice periods) 
by the inclusion of the appropriate correlation functions
$\langle H_{n_1{\bf k}-{\bf p},n_1'{\bf k}}
H_{n_2{\bf k}+{\bf p},n_2'{\bf k}}\rangle $.
We use an isotropic exponential distribution  
$f(r)=\eta^2 \exp(-r/\lambda)$ yielding 
\begin{equation} 
S({\bf p})= \eta^2 \lambda^2 
\frac{2\pi}{\left(1+(p\lambda)^2\right)^{3/2}}\,, 
\end{equation} 
where  $\eta$ denotes the standard deviation
and $\lambda$ the in-plane correlation length 
of the well-width fluctuation. It is straightforward to implement more 
sophisticated distribution functions, which might be obtained 
from Monte-Carlo simulations of the growth conditions (see, {\it e.g.,} 
Ref.~\cite{GRO95}) or X-ray characterizations of the superlattice structure  
(see, {\it e.g.,} Ref.~\cite{GRE98a}). 
Within the self-consistent Born approximation we obtain the 
self energy $\overline{\boldsymbol{\Sigma}}_{C}$ 
\begin{equation} 
\overline{\Sigma}^{</{\rm ret}}_{C;\; nn}({\bf k},E)=\sum_{\bf k'} 
\langle |H_{{\bf k}',{\bf k}}|^2\rangle  
\overline{G}^{</{\rm ret}}_{nn}({\bf k}',E) 
\end{equation} 
which provides the functional needed in the procedure scetched in 
section \ref{Secprocedure}.  
 
\section{Results}\label{Secresults} 
Let us consider the transmission of ballistic electrons through the  
superlattices considered in recent experiments by Rauch {\it et al.}  
\cite{RAU97}. 
The structure consists of $N$ wells of $6.5$ nm GaAs 
and $N+1$ barriers of $2.5$ nm Al$_{0.3}$Ga$_{0.7}$As. We obtain 
the band parameters 
$E^a=54.5$ meV, $T_1=-5.84$ meV, $K=13.25$ meV/nm  and use  
$w_{\rm eff}=10.7$ nm, where we obtained the best agreement with 
``exact'' calculations; see appendix \ref{SecAccuracy}. 
This value is somewhat larger than the well width in good agreement 
with the discussion in Section~\ref{SecCoupling}. We assume  
thickness fluctuations of half a monolayer $\eta=0.14$ nm around the 
nominal value and a correlation length  
$\lambda=5$ nm, unless otherwise stated.  
 
Motivated by the relatively sharp electron distribution injected into  
the structure, we assume that the electrons occupy the mode ${\bf k}=0$ of the  
left contact at an energy $E=E_{in}$; {\it i.e.},  
we have $f_{L{\bf k}}(E)=\delta_{{\bf k},0} 
\delta(E-E_{in})$ and $f_{R{\bf k}}(E)=0$. 
The total current through the right contact is then given by  
\begin{equation} 
J_R=\sum_{\bf k} J_{R\,{\bf k}}=-\frac{e}{\pi \hbar} 
\sum_{\bf k}\int dE\  
{\rm Tr}\left\{-i\boldsymbol{\Gamma}_{R{\bf k}}(E) \, {\bf G}^{<}(E)\right\}. 
\label{EqJGless} 
\end{equation} 
This can be expressed via Eq.~(\ref{EqBuett}) by 
\begin{equation} 
J_R=-\frac{e}{\pi \hbar} 
\sum_{\bf k} T_{(R,{\bf k})\leftarrow (L,{\bf 0})}(E_{\rm in}).  
\label{EqJtrans} 
\end{equation} 
For illustrative purpose we calculate the 
effective transmission $T(E_{\rm in})=-J_R\pi \hbar/e$ in the following. 
Regarding the applied bias we assume a homogeneous 
electric field $F$ inside the superlattice and set  
$U_L=0$, $U_n=-(n-1/2)eFd-eFb/2$ and $U_R=-NeFd-eFb=eU$ where $b$ is 
the barrier width. 
In the experiments considered, there is no charge accumulation inside  
the structure as there is on average less than one electron inside  
the structure at a given time. If necessary such effects can be 
easily taken into account by solving the Poisson equation for the  
electron density given by
\begin{equation} 
N_n=\frac{-i}{2\pi A}\sum_{\bf k}\int dE\ \overline{G}^<_{nn}({\bf k},E)\, . 
\end{equation} 
 
\begin{figure}  
\noindent 
\epsfig{file=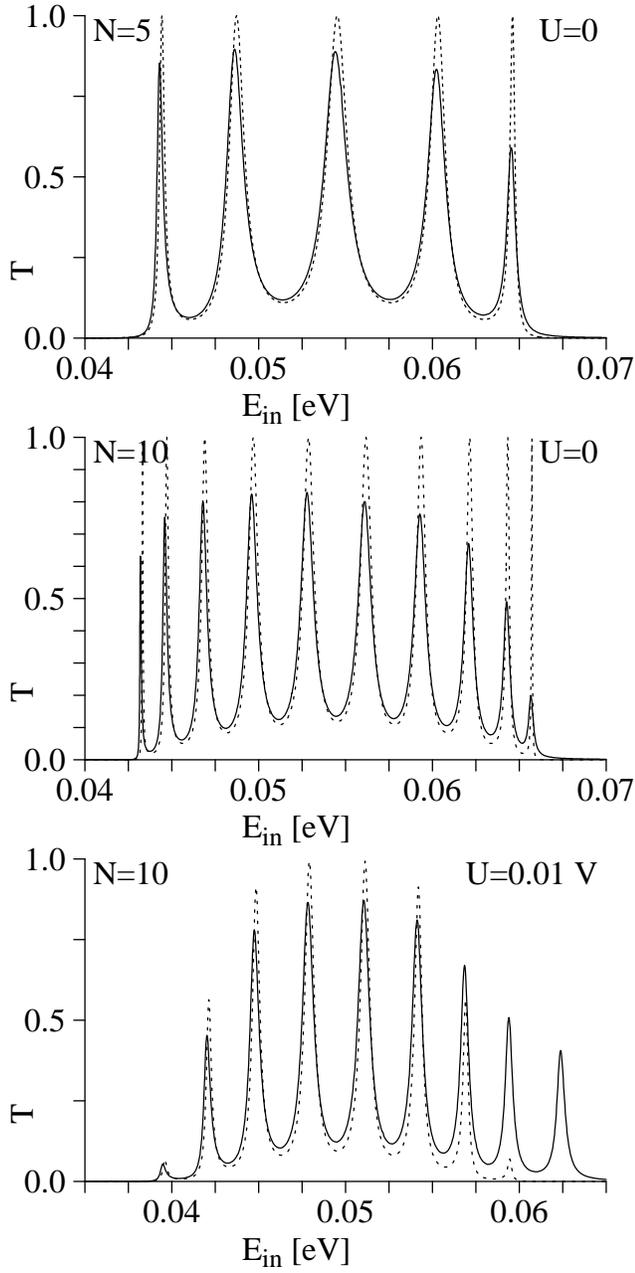,width=8.5cm}\\[0.2cm] 
\caption[a]{Transmission through  superlattices for different 
lengths and biases: The full line depicts the transmission 
calculated from Eq.~(\ref{EqJGless}) for a superlattice structure  
with interface roughness. The dotted line gives the transmission  
for an ideal superlattice without scattering.} 
\label{FigTrough} 
\end{figure}

In Fig.~\ref{FigTrough} we show the effective transmission with and without 
scattering. In both cases we find a series of peaks, equal to the number  
of quantum wells,  which reflect the eigenstates of the superlattice 
structure. For  $U= 0$  
the peak maxima reach the value 1 for the ideal  
superlattice. The broadening of these peaks results from the  
coupling to the leads and is of the order $(\Gamma_R+\Gamma_L)/N$.  
In contrast the maxima are lower and the widths are wider for the 
calculation including scattering. These effects becomes more pronounced 
with increasing superlattice length as the broadening due to scattering  
dominates with respect to the lead induced broadening.  
 
\begin{figure}  
\noindent 
\epsfig{file=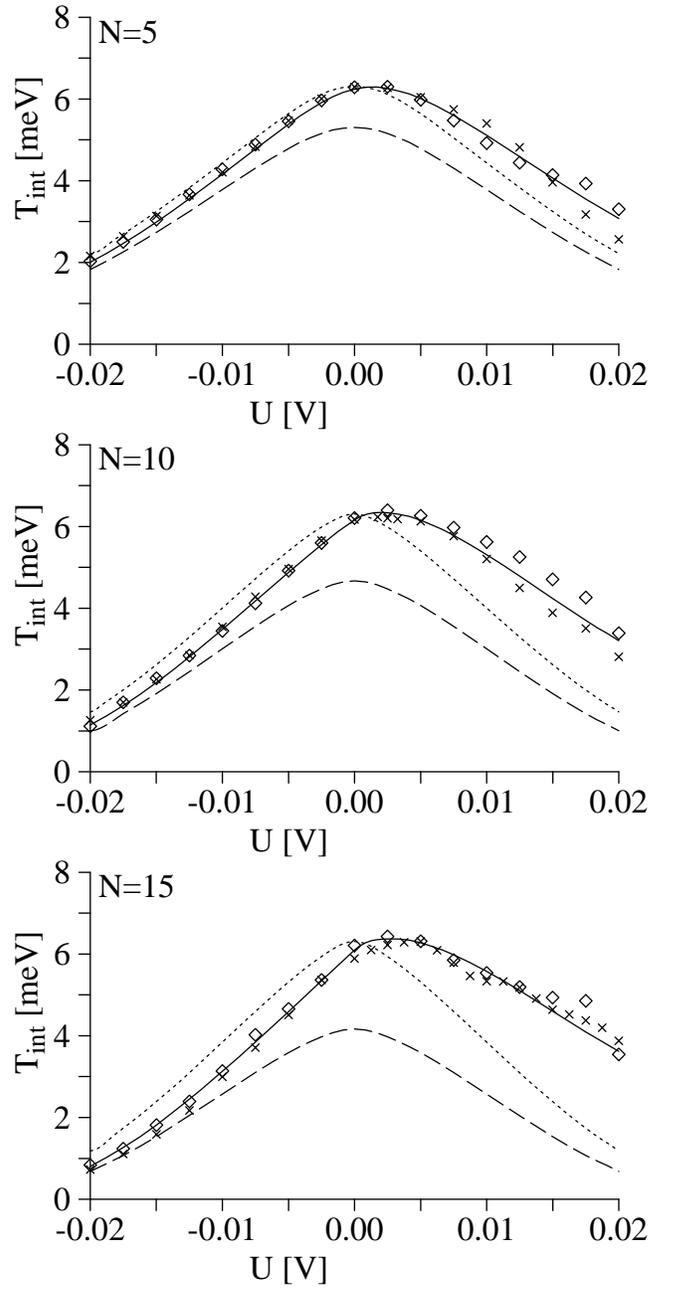,width=8.5cm}\\[0.2cm] 
\caption[a]{Integrated transmission from Eq.~(\ref{EqTint}) 
through  superlattices for different lengths: 
The full line depicts the transmission 
calculated from Eq.~(\ref{EqJGless}) for a superlattice structure  
with interface roughness. The dashed line denotes the part 
of transmission without scattering. The dotted line gives the 
result for an ideal superlattice without scattering 
for comparison. The crosses and diamonds give the results 
calculated for the same structure within the model of Ref.~\cite{WAC99a} 
for two different realizations of the interface roughness.} 
\label{FigTintrough} 
\end{figure}

An important quantity is the integrated transmission for a given  
potential drop $U$  
\begin{equation} 
T_{\rm int}(U)=\int dE_{\rm in}\ T(E_{\rm in}; U), 
\label{EqTint} 
\end{equation} 
where the integration is extended over the whole energy range of the band. 
This quantity was measured in Ref.~\cite{RAU97,RAU98}. Results are shown 
in Fig.~\ref{FigTintrough}. Let us compare the result of the  
calculations with (full line) and without  
roughness (dotted line) first.  
Without interface roughness, the function $T_{\rm int}(U)$ is always 
symmetric with respect to $U$. This can be understood from the  
symmetry property 
of the transmission matrix $T_{\ell\alpha\leftarrow \ell'\beta}(E)= 
T_{\ell'\beta \leftarrow \ell\alpha }(E)$  
(see, e.g., Ref.~\cite{DAT95}). For an ideal structure, ${\bf k}$ 
is conserved within the structure and we find  
according to Eq.~(\ref{EqJtrans}): 
\begin{eqnarray} 
T_{\rm int}(U)&=& 
\int dE_{\rm in}\sum_{\bf k}\; T_{(R,{\bf k})\leftarrow (L,{\bf 0})}(E_{\rm in}; U)  
\nonumber\\ 
&=&\int dE_{\rm in}\ T_{(R,{\bf 0})\leftarrow (L,{\bf 0})}(E_{\rm in};U)  
\nonumber\\ 
&=&\int dE_{\rm in}\ T_{(L,{\bf 0})\leftarrow (R,{\bf 0})}(E_{\rm in};U)  
\end{eqnarray} 
Now $T_{(L,{\bf 0})\leftarrow (R,{\bf 0})}(E_{in};U)= 
T_{(R,{\bf 0})\leftarrow (L,{\bf 0})}(E_{in}+eU;-U)$ due to the symmetry 
of the structure and so we find  
$T_{\rm int}(U)=T_{\rm int}(-U)$. 
 
This argument does not hold for a superlattice with interface roughness 
as the scattering is able to transfer electrons from state ${\bf k}={\bf 0}$, 
where they are injected to a finite value of ${\bf k}$. In this case 
kinetic energy $E_k$ is transferred to the $(x,y)$-direction and the electrons 
leave the superlattice with a lower $z$-component of the 
energy $E_q$. This opens up new channels 
for new processes if $U>0$; see also the discussion in Ref.~\cite{WAC99a}. 
Therefore the function $T_{int}(U)$ is asymmetric with respect to the bias $U$ 
as can be clearly seen in Figure~\ref{FigTintrough} (full line).  
These findings are in excellent agreement with recent measurements\cite{RAU98}. 
   
In Fig.~\ref{FigTintrough} we have also shown the transmission due  
to electrons traversing 
the superlattice without scattering (dashed line). This curve is obtained 
by neglecting the term $\boldsymbol{\Sigma}^{<}_{C}(E)$ 
in Eq.~(\ref{EqSigmaless}). It can be clearly seen that  
this curve is symmetric with respect to the bias and its  
magnitude is decreasing with increasing sample length. 
 
\begin{figure}  
\noindent 
\epsfig{file=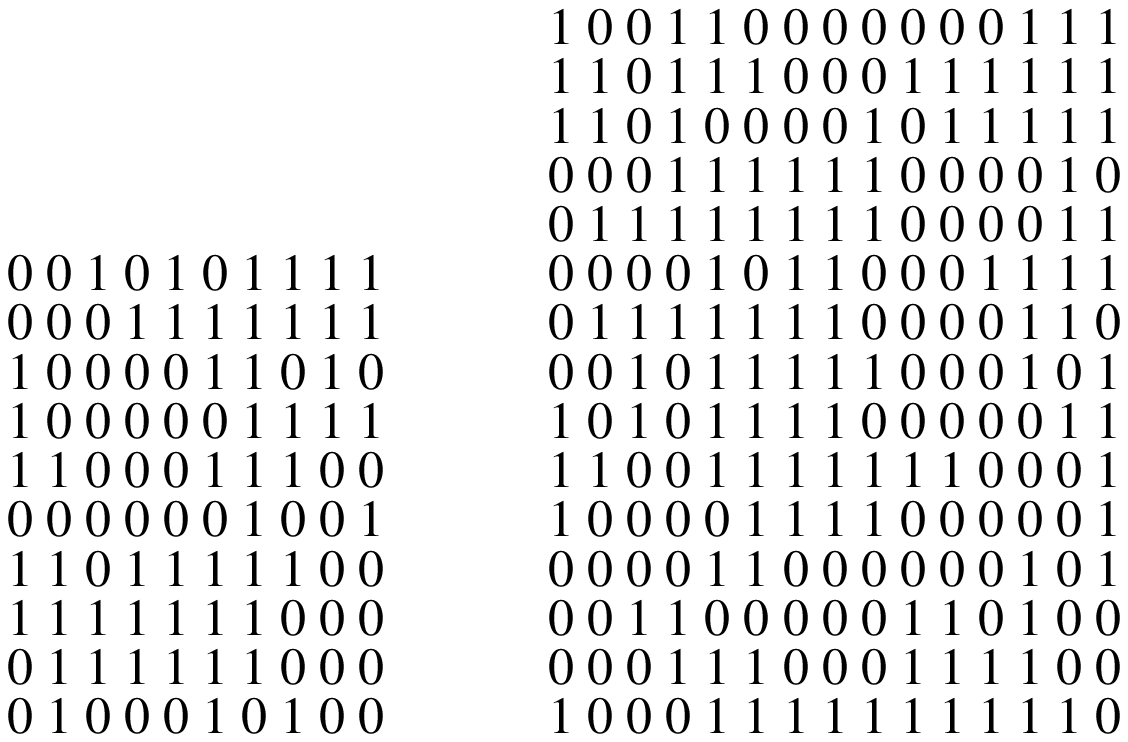,width=8.5cm}\\[0.2cm] 
\caption[a]{ Two different realizations of interface structure 
used within the model of Ref.~\cite{WAC99a}. The left one is used for  
the crosses and the right one for the diamonds in Fig.~\ref{FigTintrough}. 
Here we show the distribution of 
the second well, the other distributions have identical statistical properties. 
The spatial discretization is assumed to be 
5 nm. Both distribution give a spatial correlation 
$\langle (\xi({\bf r})-\langle \xi({\bf r})\rangle) 
(\xi({\bf r}')-\langle \xi({\bf r}')\rangle)\rangle\approx  
0.5^2\exp(-|{\bf r}-{\bf r}'|/5 {\rm nm})$ within the next 2 neighbors. 
($\xi({\bf r})$ denotes the local fluctuation.) 
} 
\label{FigInterface} 
\end{figure}

An alternative way of calculating the transmission has been performed 
in Refs.~\cite{WAC99a}. There the Green functions were calculated for 
a fixed interface potential following Ref.~\cite{TIN95}. For practical reasons 
the size of the samples is relatively small. The diamonds and crosses 
refer to two different random interface potentials as shown in  
Fig.~\ref{FigInterface} which both have approximately the 
same statistical features. The data obtained for the transmission 
are not smooth for $U>0$ and exhibit differences between each other. 
This indicates that significantly larger areas than $10\times 10$ or 
$15\times 15$ grid points must be used for reliable calculations  
utilizing this method, which is not practicable. In contrast the method using impurity 
averaging presented here gives 
a smooth behavior which, in effect, averages the scattered data points 
obtained from the previous calculations. 
 
\begin{figure}  
\noindent 
\epsfig{file=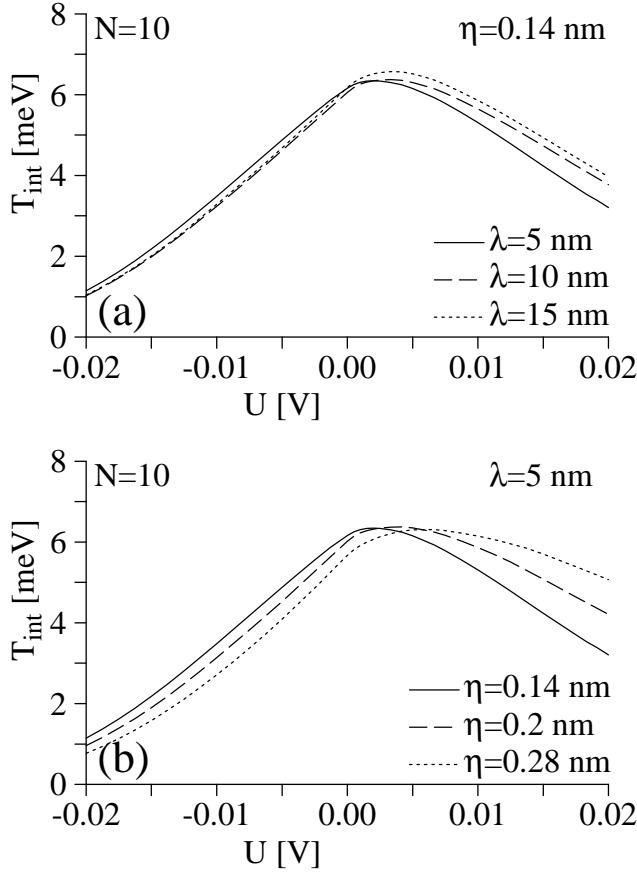,width=8.5cm}\\[0.2cm] 
\caption[a]{Integrated transmission through  superlattices with different 
correlation lengths (a) and different heights (b)  
of the interface roughness.} 
\label{Fig10ges} 
\end{figure}

In Fig.~\ref{Fig10ges}(a) we have shown the integrated transmission for different 
values of the correlation length for the roughness distributions.  
In the range considered we find that the 
asymmetry increases with the correlation length of the interface roughness. 
This indicates that larger islands lead to an enhancement of scattering 
even if the average coverage is identical. The reason is that 
scattering events with low momentum transfer is enhanced.  Such scattering 
events dominate the transport characteristics of the superlattice 
due to the energy scales involved in the system. Fig.~\ref{Fig10ges}(b) 
shows the increase of the asymmetry with the fluctuation height. 
The strong dependence allows for an estimation of the interface quality by 
analyzing the experimental transmission data.

\section{Summary and conclusions} 
We have presented a formalism to calculate the transmission 
of electrons through a finite superlattice in the presence 
of scattering processes. Due to impurity averaging  
the results are applicable to samples with large 
cross-sections.  We have also shown that  
reasonable results can be obtained by restricting the calculation 
to a basis of Wannier-functions. Within this Wannier approximation 
all couplings are well defined and can be easily calculated from 
the superlattice parameters, with the only slight ambiguity being 
the effective normalization width $w_{\rm eff}$, which is 
typically a few nanometers larger than the well width.  
 
Although we have only presented results for interface roughness 
scattering, the formalism is easily applicable to other elastic scattering 
processes, such as impurity scattering, as well. With regard to inelastic  
phonon scattering, the formalism holds as well if 
Langreth rules \cite{LAN76a,HAU96} are taken into account, which  
provide the more complicated functionals for the retarded 
and lesser self-energies, see also Ref.~\cite{WAC99b}. 
Nevertheless, one encounters the problem  
that the Green functions at different energies couple to each other. 
Therefore the set of equations which has to be solved self-consistently 
becomes significantly larger. The inclusion of electron-electron interaction 
within the mean-field model is straightforward. 
 
Our results show that interface roughness gives an enhancement of the electron 
transmission for positive biases applied to the superlattice. 
The shape of the integrated transmissions depends strongly on the distribution 
of the well width fluctuations and allows us to study interface roughness 
in semiconductor heterostructures. This provides a 
a complementary approach to the usual method of characterization 
by luminescence spectra.  
 
\acknowledgements 
Helpful discussions with S. Bose, W. Boxleitner, F. Elsholz, E. Gornik,  
A.-P. Jauho, G. Kie{\ss}lich, and C. Rauch are acknowledged. 
This work has been supported by the Deutsche Forschungsgemeinschaft in  
the framework of SFB 296. 
 
\appendix 
 
\section{Impurity averaging} \label{SecImpurityav} 
In this appendix we describe the procedure of disorder averaging  
for the superlattice  
structure discussed in Section \ref{SecApplication}. 
The cross section $A$ of typical superlattices is large   
enough that the transmission from $L$ to $R$ is given by the  
configurational average of impurities.  
For the sake of transparency we assume 
that the scattering matrix element is diagonal in the well index and  
that the impurities are uncorrelated between different wells.
The inclusion of both effects is straightforward and the
identities derived below hold in a similar way if 
$|U({\bf k}-{\bf k}')|^2$ is generalized to 
$\langle U_{n_1n'_1}({\bf k}-{\bf k}')U_{n_2n'_2}({\bf k}'-{\bf k})\rangle$.
 
\subsection{Transmission formulation} 
 
The formalism described here is similar to  
one used in Ref.\ \cite{LEO91}, except that that work was 
concerned with a resonant 
tunneling device ({\it i.e.}, one well in the structure), 
and the scattering
was calculated only to lowest order. Hence only single scattering
events were included, whereas 
the formalism described here takes multi-scattering events into 
account.  This is important for superlattices, as it is unlikely
for an electron to pass through a relatively long 
structure with only one collision. 
 
From Eqs.~(\ref{Eqtransmat}) and (\ref{EqSigmawannierk}),  
the averaged transmission matrix can be written 
\begin{eqnarray} 
\overline{T}_{R{\bf k}\leftarrow \ell_1{\bf k}_1}(E) =  
\Gamma_{R,{\bf k}}(E)\; 
\Pi(N{\bf k},\,\ell_1{\bf k}_1;E) \label{Eqtransmatimpav} 
\end{eqnarray} 
with  
\begin{equation} \begin{split}
\Pi(n{\bf k},\,\ell_1{\bf k}_1;E)  = \\
\overline{G^{\rm ret}_{n{\bf k},\,n_1(\ell_1)\, 
{\bf k}_1}(E)\,\Gamma_{\ell_1\,{\bf k}_1} 
(E)\,G^{\rm adv}_{n_1(\ell_1)\,{\bf k}_1,\,n{\bf k}}(E)}, 
\end{split}\end{equation} 
where the overline denotes impurity-averaging, $\ell_1 = L$ or $R$,   
and we define $n_1(L)=1$ and $n_1(R)=N$. 
This problem is analogous to the well-known case of impurity scattering 
in bulk material (see {\em e.g.}, Ref.~\cite{DON74}).  
The impurity averaging introduces a self-energy 
$\overline\Sigma^{\rm ret/adv}({\bf k},E)$  
to the Green functions and, as in the bulk case, vertex corrections 
due to impurity potential correlations between $G^{\rm ret}$ and 
$G^{\rm adv}$.  Within the self-consistent Born approximation, 
in addition to the impurity contribution to the self-energy,  
one must keep the ladder diagrams in the vertex\cite{DON74}.    
 
We assume that every well has the same uncorrelated randomly distributed  
concentration of impurities with areal density 
$n_{\rm imp}$, and each impurity has a Fourier transformed potential $U(q)$. 
Within the self-consistent Born approximation the retarded self-energy  
is diagonal within the well-coordinates $n,m$ [{\it i.e.},  
$\overline\Sigma_{C,nm}^{\rm ret} = \overline\Sigma_{C,n}^{\rm ret}\,\delta_{nm}$], 
and is given by  
\begin{equation} 
\overline{\Sigma}^{\rm ret}_{C,n}({\bf k},E) =  
n_{\rm imp}\int \frac{d{\bf k}'}{(2\pi)^2}\ 
|U({\bf k}-{\bf k}')|^2\, \overline{G}^{\rm ret}_{nn}({\bf k}',E). 
\label{EqSigmaretav} 
\end{equation} 
Higher order approximations have been used in Ref.~\cite{JOH92} for the 
resonant tunneling diode. 
The impurity-averaged Green function is obtained from Eq.~(\ref{EqGret}), 
and is explicitly given by  
\begin{equation}  \begin{split}
\left[{\overline{G}^{\rm ret}}^{-1}({\bf k},E)\right]_{nm} =  
\left[{{G}_0^{\rm ret}}^{-1}({\bf k},E)\right]_{nm} \\ -  
\delta_{nm}\left[\overline{\Sigma}_{C,n}^{\rm ret}({\bf k},E) 
+\Sigma_{L}^{\rm ret}({\bf k},E)\delta_{n\,1}+ 
\Sigma_{R}^{\rm ret}({\bf k},E)\delta_{n\,N}\right]\label{EqGretav2} 
\end{split}\end{equation} 
where $n,m = 1,\cdots,N$ are the well coordinates. 
The ladder approximation for  
$\Pi(n{\bf k},\,l_1{\bf k}_1;E)$ yields 
\begin{equation}  \begin{split}
\Pi & (n{\bf k},\,\ell_1{\bf k}_1;E)=\\
&\overline{G}^{\rm ret}_{n\;n_1(\ell_1)}({\bf k},E)\,\Gamma_{\ell_1{\bf k}}(E)\, 
\overline{G}^{\rm adv}_{n_1(\ell_1)\;n}({\bf k},E)\;\delta_{{\bf k}\,{\bf k}_1} \\ 
& +\sum_{m=1}^N n_{\rm imp}   \int \frac{d{\bf k'}}{(2\pi)^2}  
|U({\bf k}-{\bf k}')|^2  \overline{G}^{\rm ret}_{nm}({\bf k},E) \\
& \times \Pi(m{\bf k}',\,\ell_1{\bf k}_1;E) 
\overline{G}^{\rm adv}_{mn}({\bf k},E). \label{EqLambda} 
\end{split}\end{equation} 
This  equation can be iterated to yield $\Pi(n{\bf k},\,\ell_1{\bf k}_1;E)$.  
Note here that $\ell_1{\bf k}_1$ acts merely as a parameter.  
As we assume that particles are injected from the left side 
into the superlattice  
with zero transverse momentum, we need only calculate the case  
$\ell_1=L$ and ${\bf k}_1=0$. 
 
\subsection{Keldysh formulation} 
 
For comparison, we give below the equations 
which arise from the Keldysh formulation of this problem, 
within the same approximations 
described above. The retarded Green function 
is determined by Eqs.~(\ref{EqSigmaretav}) and (\ref{EqGretav2}), as 
in the previous subsection. 
In addition we have  
\begin{equation} 
\overline{\Sigma}^{<}_{C,n}({\bf k},E) =  
n_{\rm imp}\int \frac{d{\bf k}'}{(2\pi)^2}\ 
|U({\bf k}-{\bf k}')|^2\, \overline{G}^{<}_{nn}({\bf k}',E). 
\label{EqSigmalessav} 
\end{equation} 
Together with Eqs.~(\ref{EqKeldysh}) and (\ref{EqSigmaless}) we obtain 
\begin{equation}\begin{split} 
\Sigma^{<}_{n}({\bf k},E)  
=& i\sum_{\ell} f_{\ell{\bf k}}(E)\, {\Gamma}_{\ell{\bf k}}(E)\, 
\delta_{n\,n_1(\ell)} 
\\ 
&+ n_{\rm imp} \sum_m\int \frac{d{\bf k'}}{(2\pi)^2}\ |U({\bf k}-{\bf k}')|^2\  
\overline{G}_{nm}^{\rm ret}({\bf k}',E)\\
&\times {\Sigma}^<_m({\bf k}',E)\, 
\overline{G}_{mn}^{\rm adv}({\bf k}',E), \label{EqIterate2} 
\end{split}\end{equation} 
The numerical expense required to find the self-consistent solutions for  
$\boldsymbol{\Sigma}^{<}({\bf k},E)$ and to solve 
for Eq.~(\ref{EqLambda}) are the same. The final evaluation of  
$\overline{G}^<_{nn'}({\bf k},E)  = \sum_m  
\overline{G}^{\rm ret}_{nm}({\bf k},E)  
\Sigma^<_m({\bf k},E) \,\overline{G}^{\rm adv}_{mn'}({\bf k},E)$ 
is straightforward and we can identify  
\begin{eqnarray} 
i\sum_{\ell_1{\bf k_1}}\Pi(n{\bf k},\,\ell_1{\bf k}_1;E) 
f_{\ell_1{\bf k_1}}(E)&=& 
\overline{G}_{nn}^{<}({\bf k},E),\label{EqIdendity1} 
\end{eqnarray} 
as both sides are determined by an identical set of equations. 
Similarly, one can show 
\begin{eqnarray} 
-i\sum_{\ell_1{\bf k_1}}\Pi(n{\bf k},\,\ell_1{\bf k}_1;E)&=& 
\overline{G}_{nn}^{\rm ret}({\bf k},E)-\overline{G}_{nn}^{\rm adv}({\bf k},E) 
\label{EqIdendity2} 
\end{eqnarray} 
by using the identity  
$\overline{G}^{\rm ret}_{nn'}({\bf k},E) -  
\overline{G}^{\rm adv}_{nn'}({\bf k},E) 
=-i\sum_{m} \overline{G}^{\rm ret}_{nm}({\bf k},E)\, \Gamma_m({\bf k},E)  
\,\overline{G}^{\rm adv}_{mn'}({\bf k},E)$ where  
$\Gamma_n({\bf k},E)=i[\Sigma^{\rm ret}_n({\bf k},E) 
-\Sigma^{\rm adv}_n({\bf k},E)]$ satisfies an equation very much like  
Eq.~(\ref{EqIterate2}).  The identities Eqs.~(\ref{EqIdendity1}) and 
(\ref{EqIdendity2}) show that the Landauer-B{\"u}ttiker  
expression Eq.~(\ref{EqBuett}) with the averaged transmission  
matrix Eq.~(\ref{Eqtransmatimpav}) is identical to the Keldysh formulation 
result given by Eq.~(\ref{EqCurrentalpha}). 
This explicitly demonstrates the equivalence of the transmission and 
the Keldysh approaches within the self-consistent Born approximation of the  
scattering.

\section{The accuracy of the Wannier and wide-band approximation} 
\label{SecAccuracy} 
In order to check the accuracy of both the Wannier approximation 
and the wide-band limit, we will compare our results with  
a different approach. 
Calculations in real space 
have been performed in Refs.~\cite{TIN95,CAR71,DAT95} 
using a fine spatial discretization of length $a$.   
In the limit $a\to 0$, these calculations in principle yield  
exact results. Unfortunately, these approaches 
generate huge matrices, so that we restrict ourselves to 
a one dimensional structure and neglect the $(x,y)$-direction. 
This refers to an ideal superlattice, where the $z$ and $(x,y)$-directions 
decouple. Numbering the discretization points with indices $i$ 
the total Hamiltonian is then given by  
\begin{equation} 
H_{ij}=V_i\delta_{i,j}+\tau_{ij} 
\end{equation} 
with  
\begin{eqnarray} 
\tau_{ij}&=& 
-\frac{\hbar^2}{4a^2}\left(\frac{1}{m_{i}^*}+\frac{1}{m_{j}^*}\right) 
\qquad\mbox{for nearest neighbors},\\ 
\tau_{ii}&=&\frac{\hbar^2}{a^2m_{i}^*}, 
\end{eqnarray} 
where a position dependent effective mass $m_{i}^*$ 
has been included following 
Ref.~\cite{TIN95}.  
Now we assume that the sample is translationally  
invariant in the $z$-direction for discretization points $i< 0$ and $i> M$ 
and the coupling term in these regions is $\tau_{i,i\pm 1}=t<0$. 
Then we may define the region $0\le i\le M$ as the structure and 
the regions $i< 0$ and $i> M$ as leads within the formalism given above.   
The solutions for $j<0$ are plane waves 
$\sin(q_L(E) j a) $  and we have  
\begin{equation} 
E=U_L+2t[\cos (q_La)-1]\,.\label{EqEmode} 
\end{equation} 
For $U_L>E$ this gives an imaginary $q$ corresponding to a 
non-propagating mode.  (For a practical calculation 
$|t|$ should be larger than $|E-U_L|$;  
otherwise Eq.~(\ref{EqEmode}) does not represent the effective mass 
parabola for the leads.) 
Cutting of the leads gives a self-energy (see chapter 3.5 of Ref.~\cite{DAT95}): 
\begin{equation} 
\Sigma^{\rm ret}_{00}(E)=\left\{ 
\begin{array}{ccl} 
t\exp(i|q_L(E)a|)& \mbox{for}  & \quad E>U_L\\ 
t\exp(-|\kappa_L(E) a|)& \mbox{for}  & \quad E<U_L 
\end{array}\right. 
\label{Eqsigmadiskret} \end{equation}  
where $\kappa_L(E)$ is defined by $E=U_L+2t(\cosh \kappa_L(E)a-1)$ for $E<U_L$. 
Similar relations hold for $j=M$ with the mode from the 
right contact $R$. Note that this expression is only valid if the 
coupling from the lead to the central region is given by the same element 
$t$ as used in the discretization of the lead itself.  
The self energy is added to the potential and we obtain the matrix 
equation 
\begin{equation} 
\left[E\delta_{i,j'}-H_{ij'}-\Sigma^{\rm ret}_{ij'}\right]  
G^{\rm ret}_{j'j}=\delta_{i,j}  
\end{equation} 
which can be inverted to evaluate the Green function. 
Finally the transmission is given by the Fisher-Lee relation 
\cite{FIS81}, see also chapter 3.4 of Ref.~\cite{DAT95} 
\begin{equation} 
T_{R,L}=4t^2\, \sin \left| q_{L}(E)a\right| \, \sin \left|q_{R}(E)a\right|
\left|G^{\rm ret}_{M,0}(E)\right|^2 
\end{equation} 
which can be inserted in Eq.~(\ref{EqBuett}). 
 
\begin{figure}  
\noindent 
\epsfig{file=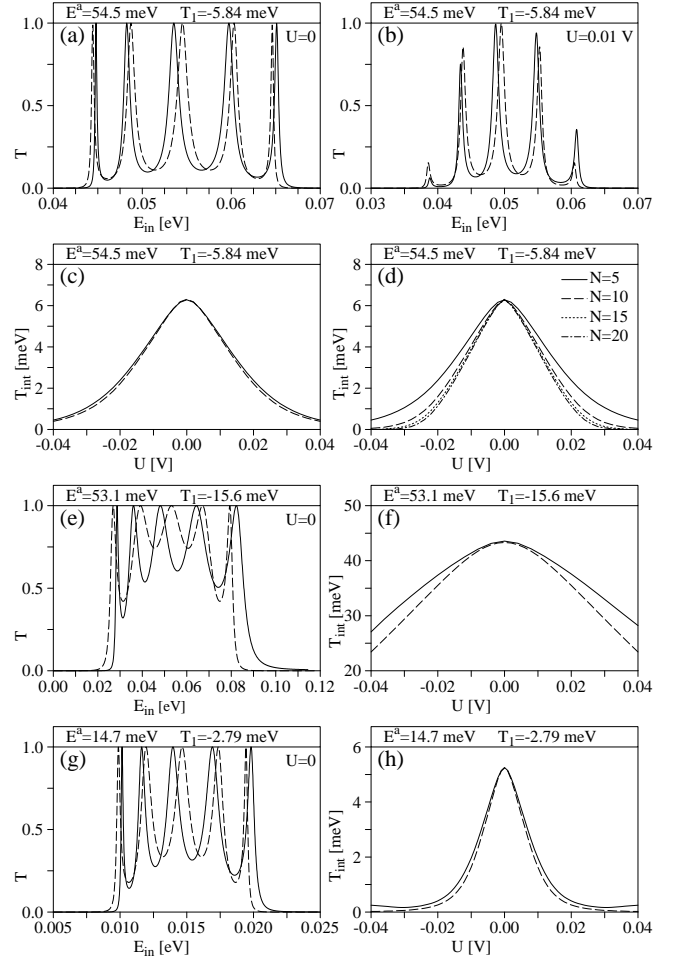,width=8.5cm}\\[0.2cm] 
\caption[a]{Transmission through ideal superlattices without 
interface roughness. (a-c) and (e-h): Comparison between the ``exact'' model  
with fine discretization (full line) and the Wannier approximation  
(dashed line) for superlattices with 5 wells. (d) Comparison of the integrated 
transmission for different superlattice lengths using  
fine discretization. The superlattice has GaAs wells with widths of 
6.5 nm (a-f) and 15 nm (g-h), and  Al$_{0.3}$Ga$_{0.7}$As barriers with 
widths of 2.5 nm (a-d) and 1 nm (e-h). The calculated parameters for  
the center of the miniband $E^a$ and the coupling $T_1$ (about a fourth of 
the miniband width) are given on top of each graph.} 
\label{FigTideal} 
\end{figure}

For comparison we consider a superlattice with N=5 wells of 6.5 nm GaAs 
and 6 barriers of 2.5 nm Al$_{0.3}$Ga$_{0.7}$As. Then we obtain 
the band parameters 
$E^a=54.5$ meV and $T_1=-5.84$ meV. The transmission is shown as a function 
of the injection energy $E_{in}$ for two different voltages in  
Figs.~\ref{FigTideal}(a) and (b).  
Here we use $w_{\rm eff}=10.7$ nm, where we obtained the best agreement. 
This value is somewhat larger than the well width in good agreement 
with the discussion in Section~\ref{SecCoupling}. 
We see that the transmission function contains 5 separate peaks which 
are related to the 5 states in the superlattice. The agreement between the 
approaches is quite good. If a bias is applied, the 
WA gives too high (low) transmissions for low (high) energies. 
The reason is the fact that the transmission through a barrier increases 
with energy, which is neglected in the WA. Preliminary results
indicate that the agreement can be improved significantly, if next-nearest
neighbor couplings are included both in Eq.~(\ref{EqGreenWannk}) and
(\ref{EqSigmawannierk}). (In this case matrix elements like
$\Sigma^{\rm ret}_{L{\bf k};\;1{\bf k},\, 2{\bf k}}(E)$
and $\Sigma^{\rm ret}_{L{\bf k};\;2{\bf k},\, 2{\bf k}}(E)$
have to be considered as well.)
 
The results for the integrated transmission 
$T_{\rm int}(U)$ are shown in Fig.~\ref{FigTideal}(c). 
We find, that the WA gives good agreement with the discrete 
model for the integrated transmission. The agreement 
becomes even better if a larger barrier width is used (not shown here). 
In Fig.~\ref{FigTideal}(d) we 
examine the length dependence of the integrated transmission calculated  
within the discrete model. We find that the function $T_{\rm int}(U)$ 
becomes almost independent of $N$ for large $N$. The results from the 
WA are almost indistinguishable and not shown here. Note,  
that $T_{int}(U)$ is a symmetric function with respect to the bias, which 
can be shown analytically using the symmetry properties of the transmission 
matrix, see Section \ref{Secresults}.  
 
In order to estimate the range of validity of the WA, we have also  
considered different superlattices. By decreasing the barrier width 
to 1 nm, we have generated a strong coupling between the wells. 
Here the miniband width is slightly larger than the center of 
the miniband.  
Best agreement between the approaches is found for $w_{\rm eff}=13.5$ nm. 
This value is larger than the one obtained above as the Wannier functions 
are less localized due to the small barrier width. 
As shown in Figs.~\ref{FigTideal}(e) and (f) the results from the 
WA deviate clearly from the ``exact'' result in this case. 
 
Finally we considered the case of a larger well width, 15 nm. 
The calculated miniband width ($4T_1$) is about  75\% of 
the the center of the miniband $E^a$. Nevertheless, the agreement 
between both approaches is still satisfactory for low biases,  
as shown in Fig.~\ref{FigTideal}(g). The second miniband 
extends from 44.8 meV to 84.7 meV in this case. Its influence can be 
seen in the integrated transmission, Fig.~\ref{FigTideal}(h). 
For $|U| > 0.024$ V the applied bias is larger than the gap between 
the lowest and second miniband. Then the coupling between the bands becomes 
important and the integrated transmission {\em increases } with bias 
for $|U| > 0.03$ V for the calculation in the discrete basis. 
Naturally this effect is not accounted for in the Wannier approximation 
due to the restriction to the lowest miniband. 
 
In conclusion we find that the Wannier approximation together 
with the wide band 
limit from Secs.~\ref{SecWA} and \ref{SecCoupling} gives good results  
if the miniband width is smaller than the  
energy of the center of the miniband and the applied bias is smaller than the 
gap between the minibands.


\fbox{ 
\begin{tabular}{c|l} 
$\ell,\ell'$ & Lead index\\ 
$\alpha,\beta$ & Subband index (in leads)\\ 
$C$ & Central region\\ 
$|\phi_{\ell\alpha q}\rangle$ & State in lead $\ell$, subband $\alpha$ with 
\\ & wavenumber $q$\\ 
$|\phi_{C,i}\rangle$ & State in central region\\ 
$i,j$ & Index of central region states\\ 
$n,m$ & Index of quantum well\\ 
$f_{\ell\alpha}(E)$ & Distribution function of lead $\ell$, subband $\alpha$\\ 
${\bf G}$ & Green function in central region (Matrix\\ &  with respect to $i,j$ or  
$n,m$)\\ 
${\boldsymbol{\Sigma}}$ & Self-energy in central region\\ 
$\boldsymbol{\Gamma}$ & $i(\boldsymbol{\Sigma}^{\rm ret} 
-\boldsymbol{\Sigma}^{\rm adv})$\\ 
$\phantom{I}^{{\rm ret},{\rm adv},<}$ &  
Retarded, advanced, ``$<$'' component of\\ &  Keldysh functions\\ 
$\overline{\bf G}$ & Impurity average (in central region) of ${\bf G}$\\ 
{\bf k} & Wavevector in $(x,y)$-plane\\ &  [{\it i.e.}, plane  $\|$ to superlattice interfaces]\\ 
$E_q$ & ``Longitudinal'' kinetic energy in lead\\ &  $\hbar^2q^2/2m$ 
(assumed $\ell,\alpha$ independent)\\ 
$E_{\ell\alpha}$ & ``Transverse" kinetic energy of lead $\ell$,\\ &  subband $\alpha$\\ 
$E^a$ & Energy of middle of miniband in central\\ &  region\\ 
$v_q$ & Longitudinal velocity of particle $\hbar q/m$ \\ & in lead\\ 
$U_\ell$ & Applied potential in lead $\ell$\\ 
$L_\ell$ & Length of lead $\ell$\\ 
$J_{\ell\alpha}$ & Current {\sl into} central region from $\ell,\alpha$\\ 
$e$ & charge of the electron ($e<0$)
\end{tabular} 
} 
\begin{center} 
Table 1.  Table of various symbols used in this paper. 
\end{center} 
 
\end{multicols}
\end{document}